\documentclass[fleqn,usenatbib]{mnras}

\usepackage[T1]{fontenc}
\usepackage[dvipsnames]{xcolor}
\usepackage[para]{threeparttable} % tables with footnotes
\usepackage{lscape}

\DeclareRobustCommand{\VAN}[3]{#2}
\let\VANthebibliography\thebibliography
\def\thebibliography{\DeclareRobustCommand{\VAN}[3]{##3}\VANthebibliography}

\usepackage{graphicx}	\usepackage{amsmath}
\usepackage{amssymb}
\newcommand{\Msun}{\,M$_\odot$}
\newcommand{\solar}[1]{\,#1$_\odot$}
\newcommand{\Rsun}{\,R$_\odot$}

\newcommand{\tento}[1]{$10^{#1}$}%tento
\newcommand{\timestento}[2]{$#1 \times 10^{#2}$}
\newcommand{\revision}[1]{{#1}}

\title[Differences in massive star models ]{Explaining the differences in massive star models from various simulations}
\author[Agrawal et al.]
{Poojan Agrawal$^{1,2,3}$\thanks{E-mail: pagrawal@astro.swin.edu.au}, 
Dorottya Sz\'ecsi,$^{4,5}$,
Simon Stevenson$^{1,2}$,
Jan J. Eldridge$^{6}$,
Jarrod Hurley$^{1,2}$
\\
$^{1}$\,Centre for Astrophysics and Supercomputing, Swinburne University of Technology, Hawthorn, VIC 3122, Australia\\
$^{2}$\,OzGrav: The ARC Centre of Excellence for Gravitational Wave Discovery, Hawthorn, VIC 3122, Australia\\
$^{3}$\,McWilliams Center for Cosmology, Department of Physics, Carnegie Mellon University, Pittsburgh, PA 15213, USA\\
$^{4}$\,Institute of Astronomy, Faculty of Physics, Astronomy and Informatics, Nicolaus Copernicus University, Grudziądzka 5, 87-100 \\ Toruń, Poland\\
$^{5}$\,I. Physikalisches Institut, Universit\"at zu K\"oln, Z\"ulpicher-Str. 77, D-50937 Cologne, Germany\\
$^{6}$\,Department of Physics, University of Auckland, Auckland, New Zealand\\
}

\date{Accepted XXX. Received YYY; in original form ZZZ}

\pubyear{2021}

\begin{document}
\label{firstpage}
\pagerange{\pageref{firstpage}--\pageref{lastpage}}
\maketitle

% Abstract of the paper
\begin{abstract}

% \dori{I edited the abstract a bit, with the aim to make it sharper and more informative. }
% \poojan{Thanks Dori.}

The evolution of massive stars is the basis of several astrophysical investigations, from predicting gravitational-wave event rates to studying star-formation and stellar populations in clusters. 
However, uncertainties in massive star evolution present a significant challenge when accounting for these models' behaviour in stellar population studies. 
In this work, we present a comparison between five published sets of stellar models from the BPASS, BoOST, Geneva, MIST and PARSEC simulations at near-solar metallicity. 
The different sets of stellar models have been computed using slightly different physical inputs in terms of mass-loss rates and internal mixing properties. 
Moreover, these models also employ various pragmatic methods to overcome the numerical difficulties that arise due to the presence of density inversions in the outer layers of stars more massive than 40\Msun{}.
These density inversions result from the combination of inefficient convection in the low-density envelopes of massive stars and the excess of radiative luminosity to the Eddington luminosity.
\revision{We find that the ionizing radiation released by the stellar populations can change by up to 18 percent, the maximum radial expansion of a star can differ between 100--1600\Rsun{}, and the mass of the stellar remnant can vary up to 20\Msun{} between the five sets of simulations. }
We conclude that any attempts to explain observations that rely on the use of models of stars more massive than 40\Msun{} should be made with caution.

\end{abstract}

% Select between one and six entries from the list of approved keywords.
% Don't make up new ones.
\begin{keywords}
stars: massive -- stars: evolution -- gravitational waves -- stars: formation -- stars: black holes -- galaxies: stellar content
%massive stars -- stellar evolution -- ionizing flux -- gravitational wave -- stellar remnant
\end{keywords}

%%%%%%%%%%%%%%%%%%%%%%%%%%%%%%%%%%%%%%%%%%%%%%%%%%

%%%%%%%%%%%%%%%%% BODY OF PAPER %%%%%%%%%%%%%%%%%%

%\LARGE

\section{Introduction}
% \vspace{-3pt}

Stellar evolutionary model sequences serve as input for a broad range of astrophysical applications; from star-formation \citep[e.g.][]{Gatto:2017} to galaxy evolution \citep[e.g.][]{Weinberger:2020}, from cluster dynamics \citep[e.g.][]{heggie_hut_2003} to gravitational-wave studies \citep[e.g.][]{VignaGomez:2018}. These sequences provide an easy and powerful way to account for both individual stars \citep[e.g.][]{Schneider:2014b} and stellar populations \citep[e.g.][]{Brott:2011b} in a given astrophysical environment.

One dimensional (1D) model sequences (from now on: \textit{stellar models}) can be computed from first principles \citep{Kippenhahn:1990} and have become a household tool in astrophysical research. 
But when it comes to stars more massive than $\sim$\,9\,M$_{\odot}$ -- those that, despite being rare, provide the bulk of the radiation, chemical pollution and the most exotic death throes in the Universe \citep{Woosley2002} -- stellar models are still riddled with large uncertainties.

High-mass stars are born less often than their low-mass counterparts \citep{Salpeter:1955} and have comparatively shorter lives \citep{Crowther2012}. Consequently, observational constraints on their evolution are more difficult to obtain.
The situation is further complicated by many massive stars being observed to be fast rotators \citep{RamirezAgudelo:2013}, which breaks down perfect symmetry, and to have a close-by companion star \citep{Sana:2012}, breaking the assumption of perfect isolation. Even for isolated, non-rotating single stars, the physical conditions both inside \citep{Heger:2000a} and around \citep{Lamers:1999} the star are so peculiar and complex that developing appropriate numerical simulations becomes highly challenging. 
This is why the evolution of massive stars remains an actively studied field to this day. 

Much progress has been made in the last few decades concerning massive stars and their evolution.
Mass loss in the form of high-velocity winds from massive stars is being intensively studied and accounted for in the models \citep{Smith:2014, Sander:2020}. %\citep[e.g.][]{Puls:2008,Surlan:2013,Smith:2014}, and accounted for in the models \citep{Bjorklund:2020,Sander:2020}. 
Observations of massive stars from the Large and Small Magellanic Clouds are being used to constrain the efficiency of interior mixing processes \citep{Brott:2011, Schootemeijer:2019}.
1D stellar models have also been updated to account for the effects of rotation \citep{Maeder2009,Costa:2019} \revision{and magnetic fields \citep{Heger:2005,Maeder:2005,Takahashi:2021}} which can significantly change their evolutionary paths \citep{Walder:2012,Petit2017,Groh2020}. 

Despite the progress, there are still many open questions surrounding the lives of massive ($\gtrsim$9\Msun{}) and `very' massive (here designated as $\gtrsim$40\Msun{}) stars, and in the absence of well-defined answers, stellar evolution codes make use of different assumptions.
Earlier studies comparing models of massive stars from different codes \citep[e.g.][]{MartinsII:2013, Jones:2015} have already established that the differences in the physical parameters such as mixing and mass-loss rates adopted by various stellar evolution codes can affect the evolutionary outcome of these stars.

Here we highlight another major uncertainty arising due to the numerical treatment of low-density envelopes of very massive stars. 
These stars have luminosities close to the Eddington-limit, so changes in the elemental opacities during their evolution can lead to the formation of density and pressure inversions in the stellar envelope \citep{Langer1997}. The presence of these density inversions can cause numerical instabilities for 1D stellar evolution codes. To deal with these instabilities, the codes use different pragmatic solutions whose interplay with mixing and mass loss can further vary the evolution of massive stars. 

% the inward gravitational force can be overcome by the outward radiation force owing to changes in the elemental opacities during the evolution of the star. When this happens, the so called Eddington-limit is said to be exceeded and density and pressure inversion regions form inside the stellar envelope. 

The role of the Eddington limit and the associated density inversions in massive stars is well known within the stellar evolution community but remains relatively unknown outside the field.
With the surge in the use of massive star models, for example, in gravitational-wave event rate predictions and supernova studies, it has become important to be aware of this issue. Our goal is to present the broader community with a concise overview, including how it affects the evolutionary properties such as the radial expansion and the remnant mass of very massive stars.
To this end, we compare models of massive and very massive stars from five published sets created with different evolutionary codes: 
\textit{(i)} models from the PAdova and TRieste Stellar Evolution Code \citep[PARSEC;][]{Bressan:2012, Cheng:2015};
\textit{(ii)} the MESA Isochrones and Stellar Tracks \citep[MIST;][]{Choi:2016MIST} from the Modules for Experiments in Stellar Astrophysics \citep[MESA;][]{Paxton2011}; 
\textit{(iii)} \revision{models \citep{Ekstroem:2012, Yusof:2013} from the Geneva code} \citep{Eggenberger:2008}; 
\textit{(iv)} models from the Binary Population and Spectral Synthesis \citep[BPASS;][]{Eldridge2017BPASS} project; and
\textit{(v)} the Bonn Optimized Stellar Tracks \citep[BoOST;][]{Szecsi:2020} from the `Bonn' Code. %To our knowledge, the present study not only compares the largest amount of different codes in the literature so far but also compares the highest stellar masses.}

% LINKS:
%PARSEC: https://academic.oup.com/mnras/article/452/1/1068/1749582
%MIST: https://iopscience.iop.org/article/10.3847/0004-637X/823/2/102/pdf
%Geneva: https://www.aanda.org/articles/aa/pdf/2012/01/aa17751-11.pdf
%BPASS: https://www.cambridge.org/core/journals/publications-of-the-astronomical-society-of-australia/article/binary-population-and-spectral-synthesis-version-21-construction-observational-verification-and-new-results/603CD23A7CF04C18EBCA253B3A1AC40C
%BoOST: https://arxiv.org/pdf/2004.08203.pdf

%METISSE: https://arxiv.org/pdf/2005.13177.pdf

We describe the major physical ingredients used in computing each set of models in Section~\ref{sec:inputs} and Section~\ref{sec:edd_lum}.
In Section~\ref{sec:comparing}, we compare the predictions from each set of models in the Hertzsprung-Russell diagram and in terms of the emitted ionizing radiation, as well as the predictions for the maximum radial expansion of stars, and their remnant masses.
Finally we draw conclusions in Section~\ref{sec:conclusions}.

% \todo{add about ionisation} \dori{Done!}
% \poojan{Thanks Dori}

\section{Physical inputs}
\label{sec:inputs}

% \begin{landscape}
\begin{table*}
	\caption{Summary of input parameters used in the computation of the models of massive stars from different codes. See Section~\ref{sec:inputs} for details. \\} 
	\centering
	\label{tab:model_proprties}
	\begin{threeparttable}
% 	\resizebox{\textwidth}{!}{%
% 	\begin{tabular}{\textwidth}{llllllll} 
	\begin{tabular}{p{0.07\textwidth}p{0.04\textwidth}p{0.09\textwidth}p{0.1\textwidth}p{0.11\textwidth}p{0.09\textwidth}p{0.04\textwidth}p{0.09\textwidth}p{0.04\textwidth}p{0.04\textwidth}}
    \hline
    \textbf{Stellar Model} & \textbf{\solar{Z}} & \textbf{Hot Wind} & \textbf{Cool Wind} & \textbf{Wolf-Rayet Wind} & \textbf{Convective boundary} &  \textbf{$\alpha_{\rm MLT}$} & \textbf{Overshoot type} & \textbf{$\alpha_{\rm ovs}$} & \textbf{$\alpha_{\rm semi}$} \\
    \hline
    \\
    BPASS & 0.020 & \citet{Vink:2000,Vink:2001} & \citet{deJagerandNieuwenhuijzen:1988} & \citet{NugisandLamers:2000} & \citet{Schwarzschild1958} & 2.0 & \citet{Pols:1998} & 0.12$^{\rm e}$  & --\\ 
    BOOST & 0.008$^{\rm a}$ & \citet{Vink:2000,Vink:2001} & \citet{NieuwenhuijzenanddeJager:1990} & \citet{Hamann:1995}$^{\rm b}$ & \citet{Ledoux1947} & 1.5 & step & 0.335  & 1.0\\ 
    GENEVA & 0.014 & \citet{Vink:2000,Vink:2001} & \citet{deJagerandNieuwenhuijzen:1988}$^{\rm c}$ & \citet{NugisandLamers:2000} & \citet{Schwarzschild1958} & 1.6$^{\rm d}$ & step & 0.1 & --\\ 
    MIST & 0.014 & \citet{Vink:2000,Vink:2001} & \citet{deJagerandNieuwenhuijzen:1988} & \citet{NugisandLamers:2000} & \citet{Ledoux1947} & 1.82 & \citet{Herwig:2000} & 0.016$^{\rm e}$ & 0.1\\ 
    PARSEC & 0.015 & \citet{Vink:2000,Vink:2001} & \citet{deJagerandNieuwenhuijzen:1988} & \citet{NugisandLamers:2000} & \citet{Schwarzschild1958} & 1.74 & \citet{Bressan1981} & 0.5$^{\rm e}$  & --\\

    	\hline
	\end{tabular}
% 	}
	
	\begin{tablenotes}
	\textbf{Notes.}
	    \item[a] For calculating mass-loss rates and opacities, \solar{Z}$=0.017$ is used.\\ 
	    \item[b] Reduced by a factor of 10. \\
	    \item[c] For ${\rm log T_{eff}/K \leq 3.7}$, mass-loss rates from \citet{Crowther2000} are used.\\
	    \item[d] For stars with initial mass $\geq$ 40\Msun{}, $\alpha_{\rm MLT}=1.0$ is used but with a different scale height (see Section~\ref{sec:edd_lum}).\\
	    \item[e] The rough equivalent in the step overshooting formalism is 0.2, 0.25 and 0.4 for the MIST, PARSEC and BPASS models respectively.
        
      \end{tablenotes}
    \end{threeparttable}
\end{table*}

\subsection{Chemical composition}

The chemical composition of the Sun is often used as a yardstick in computing the metal content of other stars. 
However, the exact value remains inconclusive and has undergone several revisions since 2004 \citep[see][for an overview]{Basu:2009, Asplund:2021}. Therefore, different stellar models often make use of different abundance scales.

The BPASS, PARSEC, Geneva and MIST models base their chemical compositions on the Sun, while the BoOST models use a mixture tailored to the sample of massive stars from the FLAMES survey \citep{Evans:2005} with $Z_{Gal}=0.0088$. For stellar winds and opacity calculations, BoOST models use $Z=0.017$ from \citet{Grevesse:1996} as the reference solar metallicity.
The BPASS models use solar abundances from \citet{Grevesse1993}
with \solar{Z}$=0.02$. The PARSEC models follow \citet{Grevesse1998} with revisions from \citet{Caffau:2011}
and \solar{Z}$=0.01524$. 
Geneva models use \citet{Asplund:2005} abundances with Ne abundance from \citet{Cunha:2006} and \solar{Z}$=0.014$  
while, finally, 
the MIST models base their abundance on \citet{Asplund:2009} with \solar{Z}$=0.01428$.

For the purpose of comparison here, we use $Z=0.014$ models for each set except for BoOST where we use the Galactic composition, $Z=0.0088$ as the closest match. 
Iron is an important contributor to metallicity, as numerous iron transition lines dominate both opacity and mass-loss rates, therefore directly affecting the structure of massive stars \citep[e.g.,][]{Puls2000}. The stellar models compared here have similar iron content, with the normalised number density ($A_{\rm Fe}= {\rm log (N_{Fe}/N_H)+12.0})$ ranging from $7.40$ (for BoOST models) to $7.54$ (for MIST models).

% The normalisation of the elemental number density NX for an element X is defined as log X ≡log (NX/NH)+12.00, for historical reasons, hence log H ≡12.00. This choice was supposedly to avoid having negative elemental abundances for the Sun (Claas 1951; Goldberg et al. 1960). While this is also true for the here recommended solar photospheric abundances – barely, with log Th = 0.03 ±0.10 being the smallest value – several naturally occurring elements indeed have log  < 0 in CI chondritic meteorites (Sect. 3): Ne, Ar, Kr, Xe, Ta, and U

% \todo{add about Fe}
% \poojan{To Jarrod: I checked the Brott et al. paper once again and found the following line in the paper: ``The resulting metallicity of our Galactic mixture is Z = 0.0088, which is lower than the solar metallicity of Z = 0.012 found by Asplund et al. (2005) and the solar neighborhood metallicity of Z = 0.014 measured by Przybilla et al. (2008).''
% For winds and opacity calculations, they use solarZ=0.017 from Grevesse et al .1996}
% \todo{add about Zsun for BoOSt models.}

\subsection{Mass-loss rates}

Stellar mass is a key determinant of a star's life and evolutionary outcome. It can, however, change as stars lose their outer layers in the form of stellar winds, and through interactions with a binary companion. 
Consequently, mass loss can  affect not only the structure and chemical composition of the star, but is also important in determining its final state \citep{Renzo:2017}. 

For massive stars, the effects of mass loss are even more pronounced.
The mass loss experienced by hot massive stars (O~type stars and Wolf--Rayet~stars) is known to be line-driven \citep{Lamers:1999} while that of cool massive stars \citep[red supergiants,][]{Levesque:2017} is suggested to be dust-driven. Both types of mass loss are an intensively studied subject. However, the complexity of the problem of atomic and molecular transitions in the wind together with the rarity of stars at these high masses means that the model assumptions are usually based either on a few observations (a small sample of stars) or on what we know about the wind properties of low-mass stars. 

All models in this study follow \citet{Vink:2000,Vink:2001} for hot wind-driven mass-loss. The PARSEC, Geneva, BPASS and MIST models follow \citet{deJagerandNieuwenhuijzen:1988} for cool dust-driven mass-loss and \citet{NugisandLamers:2000} for mass loss in the naked helium star phase.  
The Geneva models further switch to \citet{Crowther2000} for hydrogen-rich stars with ${\rm log T_{eff}/K \leq 3.7}$. They also use the maximum of \citet[][]{Vink:2001} and \citet[][]{Grafener:2008} for stars with surface hydrogen mass fraction between 0.3 and 0.05, before switching to \citet[][]{NugisandLamers:2000} when the surface hydrogen mass fraction falls below 0.05. 
The BoOST models follow \citet{NieuwenhuijzenanddeJager:1990} for cool winds and \citet[][reduced by a factor of 10]{Hamann:1995} for stars with surface hydrogen mass fraction $<\,0.3$. For computing mass-loss rates of stars with surface hydrogen mass fraction between 0.3 and 0.6, BoOST models linearly interpolate between the mass-loss rates of \citet{Vink:2001} and \citet[][reduced by a factor of 10]{Hamann:1995}.

To account for the dependence of the mass-loss rates on the chemical composition, BoOST, MIST and BPASS models scale the mass-loss rates by a factor of $Z^{0.85}$ \citep{Vink:2001} \footnote{Note that for some models the factor $Z^{0.69}$ is quoted, depending on whether the dependence of the terminal velocity on Z is explicitly considered or not.}.
Geneva and PARSEC models also use additional mass-loss as described in Section~\ref{sec:edd_lum}.

\subsection{Convection and overshooting}

Internal mixing processes such as convection and overshooting play an important role in determining both the structure and evolution of massive stars \citep[e.g., see][]{Sukhbold:2014}. 
Similar to mass loss, these processes represent another major source of uncertainty in massive stellar evolution \citep{Schootemeijer:2019, Kaiser:2020}. In 1D stellar evolution codes convection is modelled using the Mixing-Length Theory \citep[MLT]{BohmVitense1958} in terms of the mixing length parameter ${\rm \alpha_{MLT}}$. However, 3D simulations suggest that convection in massive stars might be more sophisticated and turbulent than described by MLT \citep{Jiang2015}.

The BoOST, Geneva, PARSEC and BPASS models used here follow standard MLT \citep{CoxGiuli1968} 
for convective mixing with mixing length parameter $\alpha_{\rm MLT}$\,=\,$(1.5, 1.6, 1.74, 2.0)$  respectively.
MIST follows a modified version of MLT given by \citet{Henyey1965} with $\alpha_{\rm MLT}=1.82$. 
Convective boundaries in PARSEC, Geneva and BPASS models are determined using the Schwarzschild criterion \citep{Schwarzschild1958}. BoOST and MIST use the Ledoux criterion \citep{Ledoux1947} for determining convective boundaries with semiconvective mixing parameters of 1.0 and 0.1, respectively.
For determining convective core overshoot, Geneva and BoOST use step overshooting with overshoot parameter $\alpha_{ov}$\,=\,$(0.1, 0.335)$. MIST uses exponential overshooting following \citet{Herwig:2000} with $\alpha_{ov}$\,=\,0.016. PARSEC uses overshoot from \citet{Bressan1981} with $\alpha_{ov}$\,=\,$0.5$.
BPASS uses the overshoot prescription from \citet{Pols:1998} with $\alpha_{ov}$\,=\,0.12. 
For comparison, the rough equivalent in the step overshooting formalism would be 0.2, 0.25 and 0.4 for the MIST, PARSEC and BPASS models respectively \citep[See][for details of each method.]{Choi:2016MIST,Pols:1998,Bressan:2012}

% \todo{MIST is roughly equivalent to $\delta_\mathrm{ov} = 0.2$ in the step overshooting model.}
% \todo{BPASS overshooting coefficient approximates to ${\delta_\mathrm{ov}}\sim0.4$ in the step overshooting prescription for the most massive stars ($\sim$50\Msun{}) in the set.}

MIST and PARSEC also include small amounts of overshoot associated with convective regions in the envelope. However, apart from modifying surface abundances, envelope overshoot has a negligible effect on the evolution of the star \citep{Bressan:2012}.
\revision{Rotational mixing also plays an important role in the evolution of massive stars. In fact, the calibration of the free parameters in the stellar codes is often based on their rotating models. 
For simplicity we only compare non-rotating models for PARSEC, MIST, Geneva and BPASS in this study. Although, for BoOST, in the absence of non-rotating models for stars more massive than 60\Msun{}, we do use slowly rotating (100\,km\,s$^{-1}$) models.  
As shown by \citet{Brott:2011}, this small difference in the initial rotation rate is not relevant from the point of view of the overall evolutionary behaviour.}

Major input parameters used in each set of models are summarized in Table~\ref{tab:model_proprties}.

% \begin{figure*}
% \includegraphics[width=\columnwidth,page=4]{GenevaBoOST-crop}
% \includegraphics[width=\columnwidth,page=3]{GenevaBoOST-crop}
% \includegraphics[width=\columnwidth,page=2]{GenevaBoOST-crop}
% \includegraphics[width=\columnwidth,page=1]{GenevaBoOST-crop}
% \caption{Hertzsprung--Russell diagrams of the massive single star models analysed in this work. All models have near solar composition. Symbols mark every \tento{5}\,years of evolution. Only core-hydrogen- and core-helium-burning phases are plotted; for BoOST (Bonn) models, the phase of the post-processing (cf. {Section~3} of \citealt{Szecsi:2020})
% is marked with black. Thin grey lines marks the approximate position of the observational Humphreys--Davidson limit \citep{Humphreys:1979,Gilkis:2021} where relevant. The tracks become more varied with increasing initial mass. This is because the codes apply various treatments for the numerical instabilities associated with the Eddington-limit proximity, cf. Section~\ref{sec:edd_lum}.
% }
% \label{fig:HRD}
% \end{figure*}

\begin{figure*}
\includegraphics[width=\columnwidth]{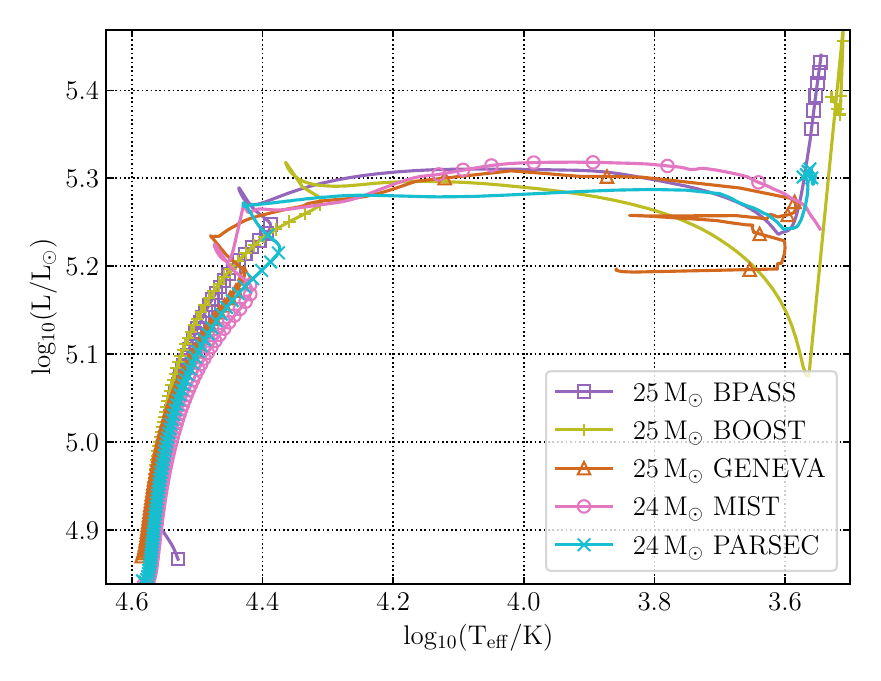}
\includegraphics[width=\columnwidth]{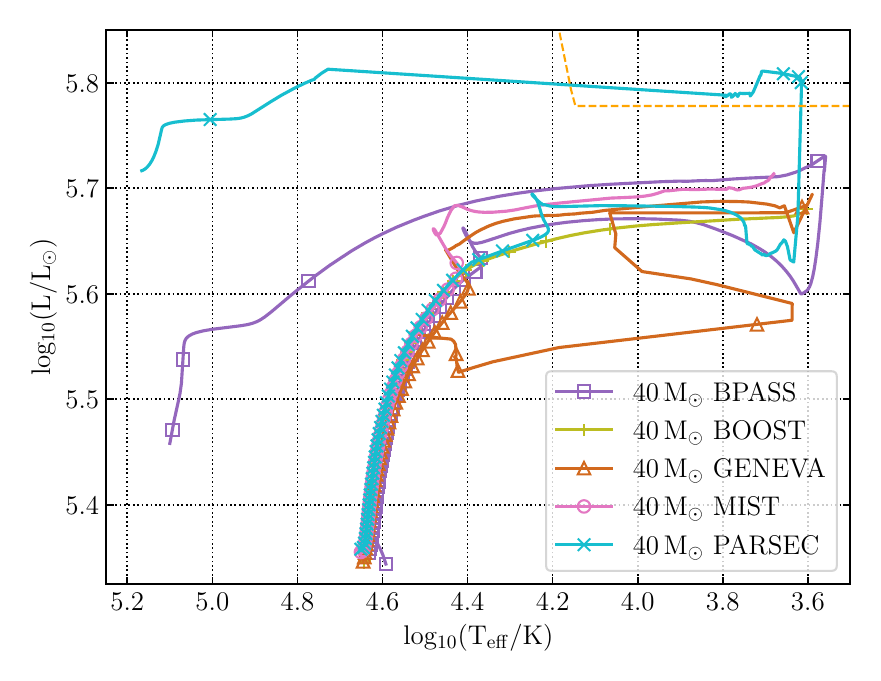}
\includegraphics[width=\columnwidth]{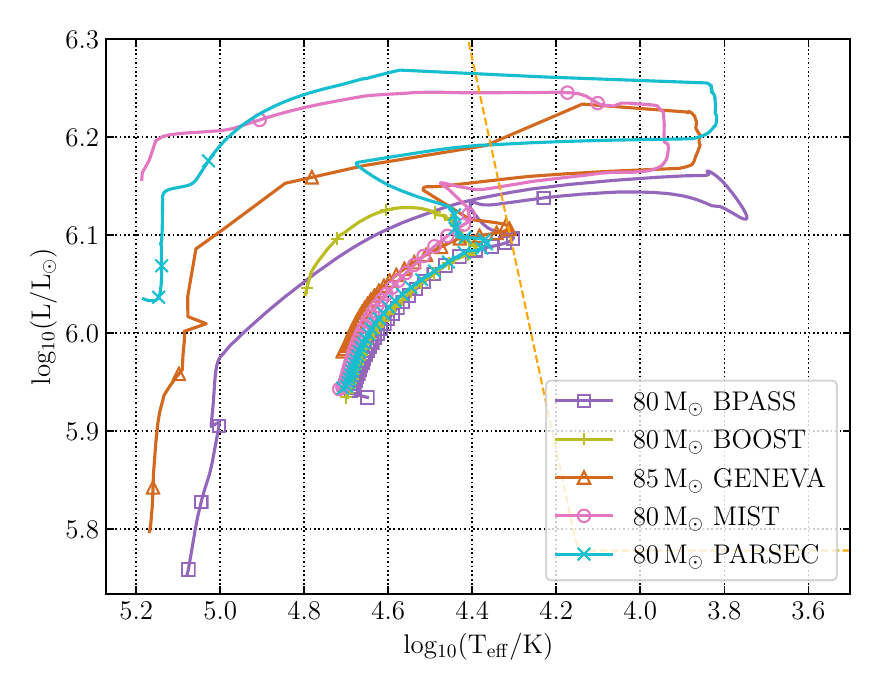}
\includegraphics[width=\columnwidth]{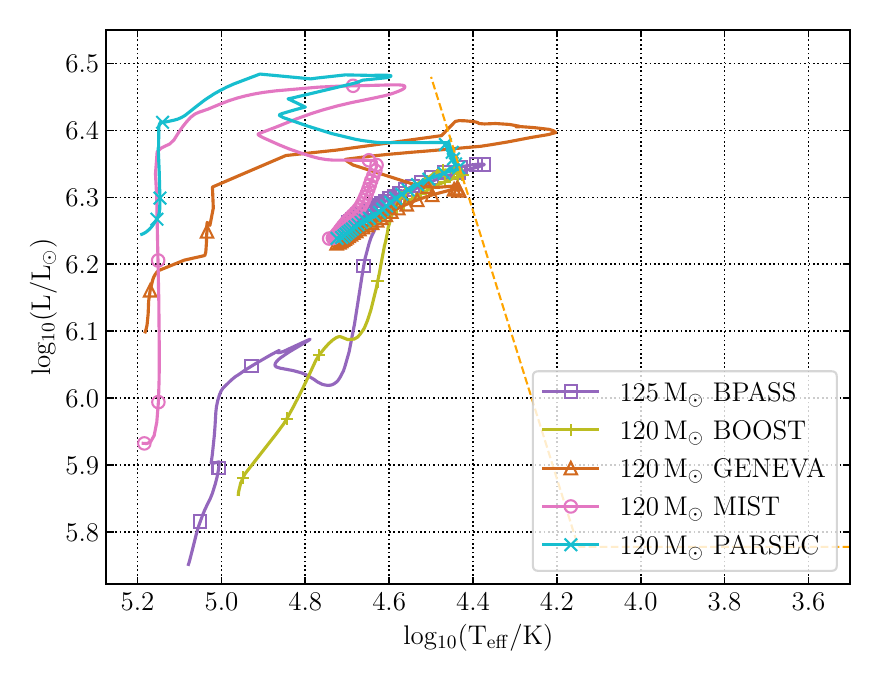}
\caption{Hertzsprung--Russell diagrams of the massive single star models analysed in this work. All models have near solar composition. Symbols mark every \tento{5}\,years of evolution. Only the core-hydrogen- and core-helium-burning phases are plotted. The dashed red line marks the observational Humphreys--Davidson limit \citep{Humphreys:1979} where relevant. The tracks become more varied with increasing initial mass. This is because the codes apply various treatments for the numerical instabilities associated with the Eddington-limit proximity, cf. Section~\ref{sec:edd_lum}.
}
\label{fig:HRD}
\end{figure*}

\section{Eddington luminosity and the numerical treatment of density inversions}
\label{sec:edd_lum}

% Massive stars have radiation dominated envelopes where the inward gravitational force is balanced by the outward radiation pressure. Under certain conditions such as the changes in the elemental opacities during the evolution of the star \citep{Cantiello:2009}, this balance can be compromised. When this happens, and the local radiative luminosity becomes greater than a critical value, the star is said to exceed the so-called Eddington-limit \citep{Langer:1997}. % (Note that this is different to the total luminosity exceeding the Eddington-limit mentioned in Section\ref{sec:initial}.)
% Eddington-limit in stars is a measure of the amount of inward gravitational force that can be overcome by the outward radiation pressure.
The Eddington luminosity is the maximum luminosity that can be transported by radiation while maintaining hydrostatic equilibrium \citep{Eddington1926}. 
In the low-density envelopes of massive stars changes in the elemental opacities during the evolution of stars can cause the local radiative luminosity to exceed the Eddington luminosity \citep{Langer1997,Sanyal:2015}.
To maintain hydrostatic equilibrium, density and pressure inversion regions form in the stellar envelope. In the absence of efficient convection (which is also typical for the low-density envelopes, \citealt{Grassitelli2016}), this can lead to convergence problems for 1D stellar evolution codes \citep{Paxton2013}. 
Owing to numerical difficulties, the time-steps become exceedingly small, preventing further evolution of the star. 
While less massive stars are only affected by this process in their late evolutionary phases \citep{Harpaz:1984, Lau2012}, very massive stars can exceed the Eddington-limit already during the core-hydrogen-burning phase \citep{Grafener:2012,Sanyal:2015} and inhibit computation of their evolution. Therefore, 1D stellar evolution codes have to employ various solutions to compute further evolution of very massive stars.

In PARSEC models, density inversions and the consequent numerical difficulties are avoided by limiting the temperature gradient such that the density gradient never becomes negative
\citep[see Sec. 2.4 of][]{Cheng:2015,Alongi1993}. Limiting the temperature gradient prevents inefficient convection and the evolution of the stars proceeds uninterrupted. 
Also, the models include a mass-loss enhancement following \citet{Vink:2011} whenever the total luminosity of the star approaches the Eddington-luminosity.

% , $\nabla_T$ to  $\frac{1-\chi_{\mu} \nabla_{\mu}}{\chi_{T}}$
%such that the density gradient, $\nabla_{\rho} \geq 0 $ 
% Thus, the most massive RSG have effective temperatures that are slightly hotter than those obtained by allowing density inversion to occur.
% This is again a technical solution that may or may not hold in real stars. 

% Various codes deal with these challenges in various ways. One common solution is to artificially enhance the mass-loss rate, thereby removing layers from the surface which helps the models to converge. This artificial enhancement of the mass loss (well above values predicted by the standard prescriptions listed in Section\,\ref{sec:initial})} is a pragmatic method to get rid of numerically unstable layers. 

MIST models suppress density inversions through the MLT++ formalism \citep{Paxton2013} of MESA. In this method, the actual temperature gradient is artificially reduced to make it closer to the adiabatic temperature gradient whenever radiative luminosity exceeds the Eddington luminosity above a pre-defined threshold. This approach again increases convective efficiency, helping stars to overcome density inversions.
Additionally, radiative pressure at the surface of the star is also enhanced in the MIST models to help with convergence \citep{Choi:2016MIST}.

In the extended envelopes of massive stars, the density scale height is much larger compared to the pressure scale height (which is typically used for computing the mixing length). Therefore, setting the mixing length to be comparable with the density scale height helps avoid density inversion \citep{Nishida1967, Maeder1987}. 
The Geneva models include this treatment when computing models with initial masses greater than 40\Msun{} with $\alpha_{\rm MLT}=1.0$ \citep[see Sec. 2.3 of][]{Ekstroem:2012}. Additionally, the mass-loss rates for the models are increased by a factor of three whenever the local luminosity in any of the layers of the envelope is higher than five times the local Eddington luminosity.  

% The side effect of this treatment is that the redward extension of the tracks in the Hertzsprung- Russell (HR) diagram is reduced by 0.1-0.2 dex in Teff (see Maeder & Meynet 1987),

BoOST models do not include any artificial treatment to prevent massive stars from encountering density inversions. Instead, their models undergo envelope inflation when massive stars reach the Eddington limit \citep{Sanyal:2015}. On encountering the density inversions in their envelopes, the computation of very massive stars becomes numerically difficult. Further evolution of such stars is then computed through post-processing. It involves removing layers from the surface of the star (which would anyway happen due to regular mass loss) while correcting for surface properties such as effective temperature and luminosity \citep[][]{Szecsi:2020}.

BPASS models also allow density inversions to develop in the envelope of massive stars. However, these models are able to continue the evolution without numerical difficulties, most likely due to the use of a non-Lagrangian mesh \citep[see][for an overview]{Stancliffe:2006} and the resolution factors being lower than in other models \citep[][]{Eldridge2017BPASS, Eggleton1973}.

\section{Comparing the models}
\label{sec:comparing}

\subsection{Differences between models in the Hertzsprung--Russell diagram}
\label{sec:HRD}

The evolution of stars can be easily represented through tracks on the Hertzsprung-Russell (HR) diagram, depicting the evolutionary paths followed by a series of stars. Figure~\ref{fig:HRD} presents the HR diagram of stars of various initial masses from the five simulation approaches. The observational analogue to the Eddington-limit is the Humphreys–Davidson limit or HD-limit \citep{Humphreys:1979}. 
Since the luminosity of a star depends on its mass, more massive stars also are more luminous. This means they can easily exceed the Eddington-limit, develop density inversions and require the use of numerical solutions as discussed in Section~\ref{sec:edd_lum}.

From Figure~\ref{fig:HRD} 
we see that the tracks of the 25\Msun{} (or 24\Msun{} in some cases) 
stars agree well during most of the evolution. 
This is because stars of this mass do not exceed the Eddington limit 
and are thus not affected by the related numerical treatments. 
%they are not affected by the numerical treatments for proximity to the Eddington-limit. 
%Therefore, their tracks agree during most of the evolution. 
The minor differences in their tracks are due to the difference in physical inputs (Section~\ref{sec:inputs}) between the simulations. 
For example, the differences in the position of the main-sequence (MS) hook feature in the HR diagram arise due to the varied extent of convective overshoot used in each set of models.

% . The other difference arises in the Hayashi limit for each model, which again results from the differing amount of convective mixing. 

% This feature depends on the various internal mixing parameters applied by the codes; we refer to e.g. \citet{Castro:2018} for the discussion of overshooting, and to \citet{Schootemeijer:2019} for that of semi-convection. But for very massive stars, these factors become negligible compared to the overwhelming effect of mass loss. 

% For example, main-sequence's endpoint is highly sensitive to the amount of overshooting \citep{Castro:2014,Castro:2018} while core-helium-burning is strongly influenced by the assumed efficiency of semi-convection \citep{Schootemeijer:2019}, amongst other things \citep{Gilkis:2021}. 

A 40\Msun{} star is clearly affected by the numerical treatment employed during the post-main sequence phase of its evolution, 
as evidenced by the difference in the tracks in the HR diagram 
shown in Figure~\ref{fig:HRD}. 
More massive stars, i.e, those with initial masses 80/85\Msun{} and 120/125 \Msun{}, can exceed the Eddington-limit in their envelopes while on the MS and therefore their simulations differ significantly from each other. 
At these masses, the mass-loss rates can be as high as  10$^{-3}$\,$-$\,10$^{-4}$\,M$_{\odot}$\,yr$^{-1}$, completely dominating over every other physical ingredient in determining the evolutionary path.
While all tracks have been computed with similar prescriptions for wind mass loss (cf. Section~\ref{sec:inputs}), the actual rates can be  strongly modified by the numerical methods adopted by each code in response to numerical instabilities (Section~\ref{sec:edd_lum}), resulting in vast differences in the tracks.

% The second main reason why the most massive models are so varied is related to how the modellers deal with the uncertainties related to the Eddington-limit proximity of the models; we explain this in Section~\ref{sec:reasons}.

\begin{table*}
	\centering
	\caption{\revision{Time averaged ionizing photon number flux [s$^{-1}$] in the Lyman continuum emitted by the stellar models during their lives \textit{on average}, cf.~Section~\ref{sec:ion}.} The last column provides the amount of Lyman radiation (number of photons [s$^{-1}$]) that a 10$^7$\Msun{} population (e.g. a starburst galaxy or a young massive cluster in the Milky Way) containing these massive stars would emit.
% 	\poojan{To Dori: Is formatting of this table correct?}	\dori{I think so? Why?}\poojan{Sorry, I should have been more explicit. It's how the numbers are written in the table e.g., 4.5e49 versus \timestento{4.5}{49}. See below and comment out the one you don't like.}	\simon{I think I prefer the second one (\timestento{4.5}{49})}
	} 
	\label{tab:ion}
	\begin{tabular}{lccccc} 
		\hline M$_{\rm ini}$ [M$_{\odot}$]
		& 24/25 & 40 & 80/85 & 120/125 & pop. \\
		\hline  
% 		PARSEC  & 3.7e48 & 1.3e49 & 5.5e49 & 1.0e50 & 1.08e54\\ 
% 		MIST  & 3.3e48 & 1.5e49 & 5.1e49 & 1.1e50 & 1.06e54\\ 
% 		Geneva  & 3.5e48 & 1.2e49 & 4.6e49 & 7.8e49 & 9.27e53\\ 
% 		BPASS  & 3.6e48 & 1.3e49 & 4.5e49 & 7.7e49 & 9.34e53 \\ 
% 		BoOST & 3.7e48 & 1.2e49 & 4.4e49 & 7.4e49 & 9.14e53\\ \hline
		
		PARSEC  & \timestento{3.7}{48} & \timestento{1.3}{49} & \timestento{5.5}{49} & \timestento{1.0}{50} & \timestento{1.08}{54}\\ 
		MIST  & \timestento{3.3}{48} & \timestento{1.5}{49} & \timestento{5.1}{49} & \timestento{1.1}{50} & \timestento{1.06}{54}\\ 
		Geneva  & \timestento{3.5}{48} & \timestento{1.2}{49} & \timestento{5.1}{49} & \timestento{8.5}{49} & \timestento{9.90}{53}\\ 
		BPASS  & \timestento{3.6}{48} & \timestento{1.3}{49} & \timestento{4.5}{49} & \timestento{7.7}{49} & \timestento{9.34}{53} \\ 
		BoOST & \timestento{3.7}{48} & \timestento{1.2}{49} & \timestento{4.2}{49} & \timestento{6.9}{49} & \timestento{8.89}{53}\\ 
		
		\hline
	\end{tabular}
\end{table*}

\subsection{Ionizing radiation and synthetic populations}
\label{sec:ion}

% \dori{I have corrected this section. There were some outdated lines in it. I also edited the explanations a little bit, for clarity. I believe it is now correct.}
% \poojan{Looks great!!}

The ionization released by a stellar population in e.g. a cluster or galaxy is influenced by the contribution of the most massive stars \citep{Topping:2015}.
As shown above however, these are the stars for which the simulations give the most diverse predictions. 

To demonstrate this effect, we calculate the ionizing radiation emitted by a simple stellar population, supposing a \citet{Salpeter:1955} initial mass function \revision{with an upper mass of 120\Msun{}} and a star-forming region of 10$^7$\Msun{} total mass -- which is aimed to represent either a typical starburst galaxy, or a young massive cluster in the Milky Way. In Table~\ref{tab:ion} we list the Lyman photon flux predicted by the individual stellar models analysed here. \revision{To simplify the population synthesis calculations, the table provides time averaged values, that is, the photon number flux emitted over the whole evolution is divided by the lifetime. This way the emission coming from the population is estimated by simply weighting the time averaged values by the initial mass function (i.e. without needing to follow the time evolution of the modelled cluster or galaxy). The results of these simple population syntheses are also reported in Tab.~\ref{tab:ion}.}
In the absence of spectral synthesis models computed for all five sets \citep[cf.][]{Wofford:2016}, we have opted to simply use black body estimation. To correct for optically thick winds, we follow the method explained in Chapter~4.5.1 of \citet[][which relies on \citealt{Langer:1989a}]{Szecsi:2016}. 

\revision{We find that, in terms of how much Lyman flux is emitted by a given synthetic population, the model predictions can differ as much as $\sim$18 percent between simulations.}
This supports earlier findings \citep[e.g.][]{Topping:2015} that relying on the ionizing properties of massive stars from evolutionary models should be done with caution. Indeed one should keep in mind that the behaviour of the most massive models, those that dominate the radiation profile of any star-forming region, is weighted with large uncertainties -- the source of which is the treatment of the Eddington limit, explained in Section~\ref{sec:edd_lum}.

\subsection{Predictions of maximum stellar radii}
\label{sec:maxrad}

The radial expansion of a star plays a significant role in determining the nature of binary interaction as it can lead to episodes of mass transfer in close interacting binaries. Recent studies indicate that most of the massive stars occur in binaries \citep{Sana:2012,Moe:2017}.
Therefore, predictions of stellar radii become even more important for determining the binary properties of massive stars. 

\begin{figure}
\includegraphics[width=\columnwidth]{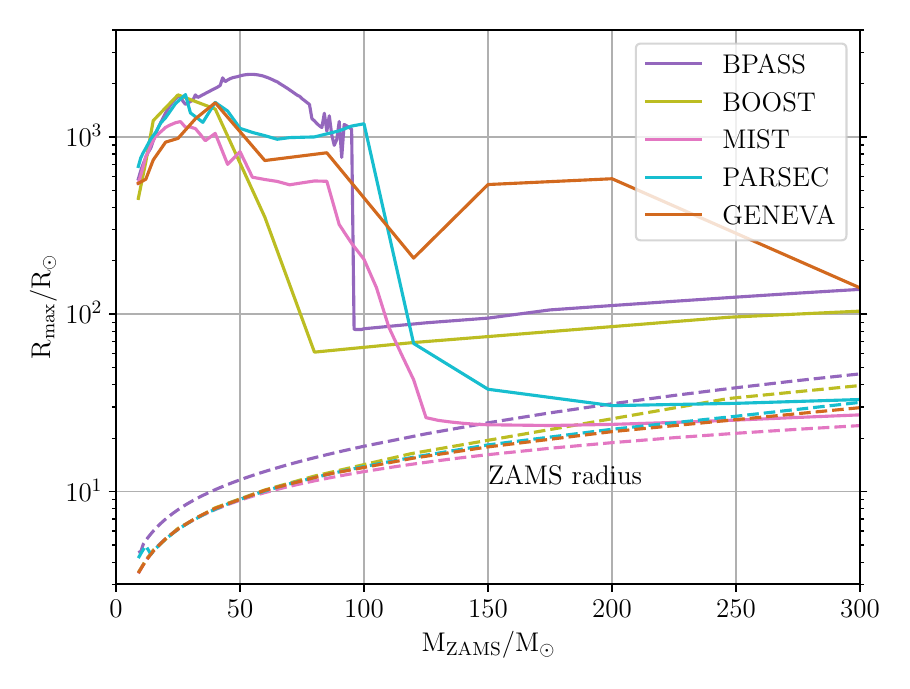}
\caption{Maximum stellar radii as a function of the initial mass of the star. Similar to Figure~\ref{fig:HRD}, differences in the physical inputs and the numerical methods adopted by each code can lead to a difference of more than $1000$\Rsun{} in predictions in terms of the maximum radial expansion achieved by the stars.}
\label{fig:maxrad}
\end{figure}

Figure~\ref{fig:maxrad} shows the maximum radial expansion achieved by massive stars from each simulation. For stars with initial mass up to 30\Msun{}, all simulations predict the formation of a red supergiant. The maximum difference in the radius predictions here is $\lesssim$1000\Rsun{}. For higher initial masses, the predictions for maximum stellar radii become more divergent as proximity to the Eddington-limit increases and numerical treatments adopted by each code modify the mass-loss rates. 

The greatest difference in the maximum radius predictions ($\gtrsim$1000\Rsun{}) occurs for stars with initial masses between 40 and 100\Msun{}. Above $\sim$\,100\Msun{}, stars have even higher mass-loss rates which can completely strip a star of its envelope before it can become a red-supergiant. Such stars evolve directly towards the naked helium star phase and have much smaller radii. 
Therefore, for stars with initial masses more than 100\Msun{}, the difference between the maximum radius predictions by each simulation reduces to $\lesssim$100\Rsun{}. 
The predictions in this mass range seem to further converge into two main groups: PARSEC and MIST represent one group predicting smaller radii compared to the second group which consists of BoOST and BPASS models. 
\revision{The maximum radii in this mass range are predicted by the Geneva models. This is due to the difference in the mass loss rates adopted during the naked helium phase of the star, as explained in the next section.}
% Since the maximum initial mass of a star in the Geneva set is 120\Msun{}, we cannot comment on the behaviour of their models outside of this range.

\subsection{Remnant mass predictions}
\label{sec:remnant}

\begin{figure}
\begin{tabular}{c}
\includegraphics[width=\columnwidth]{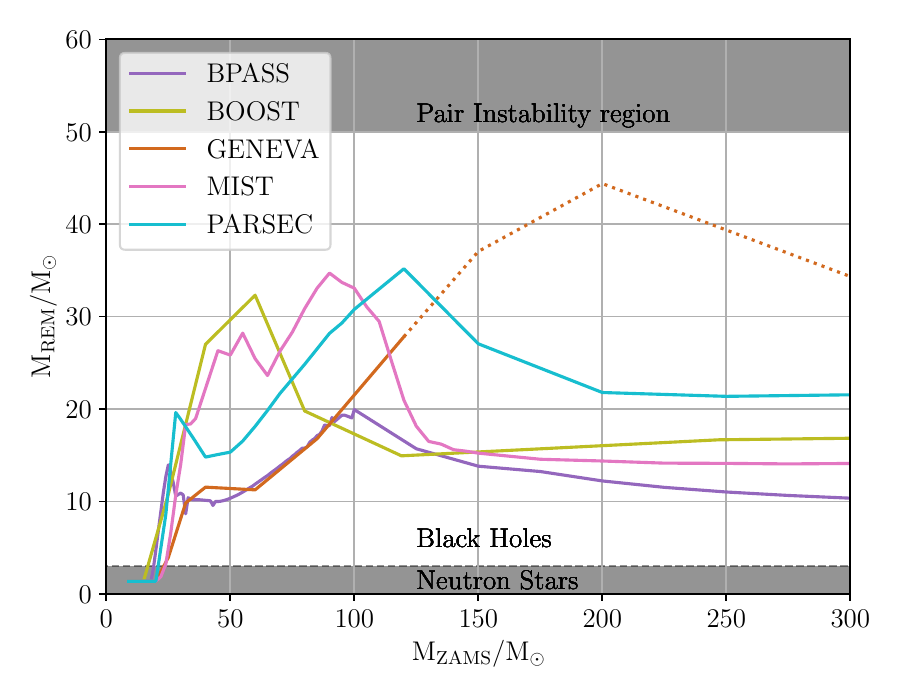}
\\
\vspace{-10pt}
\includegraphics[width=\columnwidth]{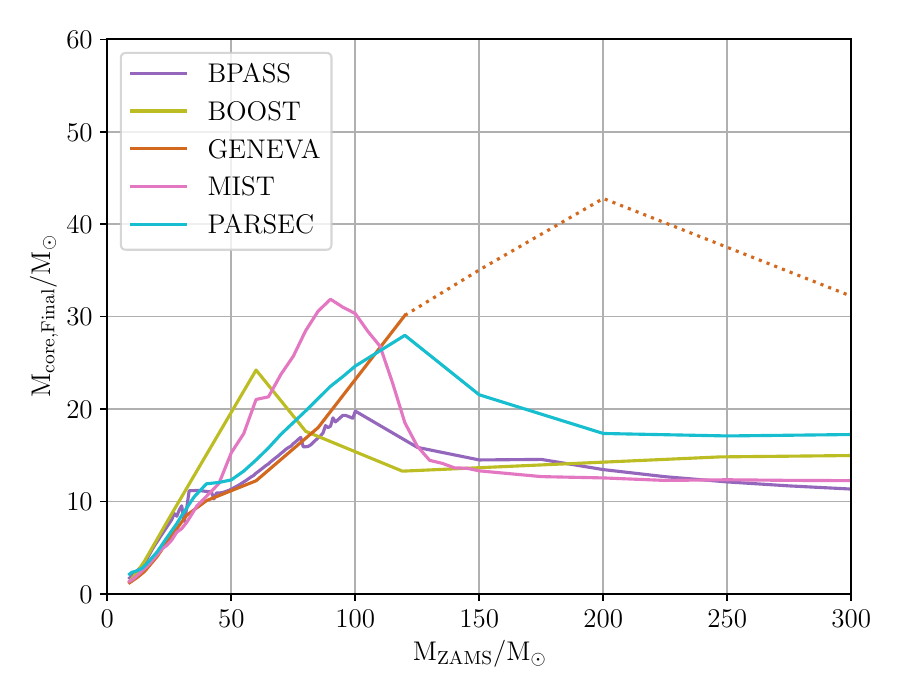}
\\
\vspace{-10pt}
\includegraphics[width=\columnwidth]{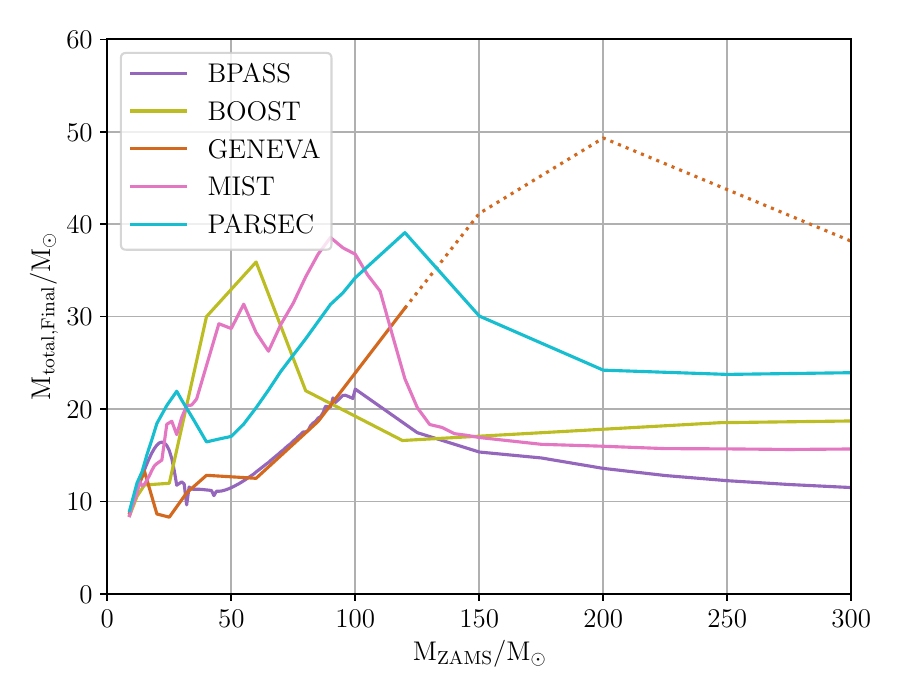} 
\end{tabular}

\caption{Final masses of stars as a function of their initial mass, ${\rm M_{ZAMS}}$. The top panel shows the mass of stellar remnants as predicted by the different sets of stellar models. The middle panel shows the carbon-oxygen core mass and the bottom panel shows the total mass of the star, as used in the calculation of the remnant masses. \revision{For stars more massive than 120\Msun{}, the Geneva models from \citet{Yusof:2013} use a different criteria for defining CO-core compared to the lower mass models from the same set (see Section~\ref{sec:remnant} for details) and therefore, are represented using dotted line.} Differences in the evolutionary parameters for massive stars can cause variations of about 20\Msun{} in the remnant masses between the stellar models from various simulations.}

\label{fig:remnant}
\end{figure}

% \caption{Mass of the stellar remnant as a function of the initial mass of the star. Since all the models we study have near-solar metallicity and therefore rather high mass-loss rates, none of them stays massive enough at the end of their lives to undergo pair-instability (the Geneva models do not provide information on core masses, hence we omit them from this analysis). Differences in the assumptions in massive star modelling can cause a variation of up to 20\Msun{} in the remnant masses between simulations. }
% Thus, choosing to apply one of these simulations over the others in e.g. gravitational-wave event rate predictions can lead to strikingly different results.}

Stellar evolutionary models provide an easy way of estimating the properties of stellar remnants such as black holes and neutron stars, which are needed in many fields including supernova studies \citep[e.g.][]{AguileraDena:2018, Raithel:2018}, gamma-ray bursts progenitors \citep[e.g.][]{Yoon:2006, Szecsi:2017long},  and gravitational-wave event rate predictions \citep[e.g.][]{Stevenson:2019, Mapelli:2020}.

Following \citet[][]{Belczynski:2010} we show in Figure~\ref{fig:remnant} how the uncertainties in the models we compare here also pose a challenge for the predictions of remnant properties.  
Remnant masses have been calculated from the carbon-oxygen (CO) core mass and the total mass of the star at the end of the core helium-burning phase using the prescription of \citet{Belczynski:2008} \citep[same as the StarTrack prescription in][]{Fryer:2012}.

\revision{The Geneva models do not provide information on core masses in their publicly available dataset. Therefore, the final CO-core mass for these models has been taken from \citet[][for stars with initial mass up to 120\Msun{}]{Georgy:2012} and \citet[][for stars with initial mass more than 120\Msun{}]{Yusof:2013}.
Note that in the Geneva models from \citet{Yusof:2013}, CO-core mass is defined as the mass of the core where the sum of mass fraction of carbon and oxygen exceeds 75 percent, and is different to the definition used by \citet{Georgy:2012}. } 

% These models have been computed using the same code with the same physical ingredients as the other Geneva models from \citet{Ekstroem:2012}.

\revision{The remnant masses are heavily influenced by the modelling assumptions (cf. Section~\ref{sec:inputs}) and the numerical methods (cf. Section~\ref{sec:edd_lum}) especially above M$_{\rm ini}$~$=$~40\,M$_{\odot}$ where the most massive black holes are predicted. For stars with initial masses between 9 and 120\Msun{}, we find that the mass of the black holes predicted by the different sets of models can differ by $\sim$\,20\Msun{}. The maximum black hole mass varies from about 20\Msun{} for BPASS models to about 35\Msun{} for MIST and PARSEC models, and 32\Msun{} for the BoOST models. BPASS models consistently predict the lowest values of remnant mass for most of the massive stars while predictions from the BoOST, MIST and PARSEC models peak at 60, 90 and 120\Msun{} before flattening out at the higher initial masses.}

\revision{For more massive stars, the variation in the remnant mass between BPASS, BoOST, MIST and PARSEC models reduces to about 10\Msun{}. 
An interesting behaviour is shown by the Geneva models, which predict one of the lowest remnant masses for stars up to 120\Msun, reaching a maximum of only 28\Msun{} at 120\Msun{}. These models however, predict the highest values of remnant masses beyond 120\Msun{}. At their farthest, for a model with initial mass of 200\Msun{}, the predictions between the Geneva models and other models can be as high as 30\Msun{}. 
Similar variability is found in the core mass and the final total mass of the star (from which the remnant masses have been calculated). }

% This definition leads to a prediction of smaller core  compared to core masses predicted by \citet[][]{Georgy:2012} for the same mass models. However, according to \citet{Belczynski:2008} method used here, stars with c are predicted to directly collpase into black hole at the end of their lives, thus 

\revision{The significantly higher remnant masses predicted by the Geneva models for stars with initial mass beyond 120\Msun can be explained as follows.
Due to their high luminosity, stars more massive than 100--120\Msun{} rapidly lose mass during the main-sequence phase and directly evolve towards the naked helium phase (cf. Figure~\ref{fig:HRD}). 
The mass-loss prescriptions from both \citet[][]{NugisandLamers:2000} and \citet[][]{Hamann:1995} predict the highest mass-loss rates for stars occur during this naked helium phase. 
BPASS, MIST and PARSEC models switch to mass-loss rates from \citet[][]{NugisandLamers:2000} when the mass fraction of hydrogen at the surface of the star falls below 0.4, 0.4 and 0.5 respectively. BoOST models linearly interpolate between the mass-loss rates of \citet{Vink:2001} and \citet[][reduced by a factor of 10]{Hamann:1995} for stars with surface hydrogen mass fraction between 0.3 and 0.6, before completely switching to the mass-loss rate from \citet[][reduced by a factor of 10]{Hamann:1995} for stars with surface hydrogen mass fraction less than 0.3.}

\revision{Geneva models, on the other hand, switch to using mass-loss rates from \citet[][]{NugisandLamers:2000} only when the mass fraction of hydrogen at the surface of the star falls below 0.05. For stars with surface hydrogen mass fraction between 0.3 and 0.05, they use the maximum of \citet[][]{Vink:2001} and \citet[][]{Grafener:2008}, which predict lower mass-loss rates compared to both \citet[][]{NugisandLamers:2000} and \citet[][]{Hamann:1995} \citep[See section 2.1 of][]{Yusof:2013}. Thus, they do not lose as much mass as other models during the naked helium star phase and end with higher total mass and thus with the higher remnant mass.}

\revision{Note that the \citet{Belczynski:2008} prescription is one of several methods for predicting the remnant properties of the stars.
Other methods for calculating the remnant masses may predict higher or lower values. For example, the remnant mass calculations based on the binding energy of the star \citep[e.g. see][]{Eldridge2017BPASS} are generally lower than those predicted here. However, all the models we study have near-solar metallicity and therefore rather high mass-loss rates, none of them stays massive enough at the end of their lives to undergo pair-instability. }

% The Geneva models do not provide information on core masses, hence we omit them from the analysis of the core mass and the remnant mass. However, we do show their predictions for the final total mass which,  for stars up to 100\Msun{}, are similar to the predictions of BPASS models. 

% Similar to the predictions of maximum stellar radii in Figure~\ref{fig:maxrad}, the predictions of final masses for stars with initial mass $\gtrsim$\,100\Msun{} separate into two groups, although the groups are slightly different this time, BPASS and MIST represent one group predicting smaller final masses compared to the second group consisting of BoOST and PARSEC models.

% These differences indicate the sensitivity of stellar models to physical inputs. Since very massive stars have convective cores,  their evolution is dominated by the assumed mixing parameters, especially the amount of convective overshoot. On the other hand, evolution of the total mass depends critically on mass-loss rates.

% \subsection{Discussion of the differences}
% \label{sec:reasons}

% Our understanding of interaction due to a binary companion is growing rapidly \citep{deMink:2014,Vanbeveren:2018,Shenar:2019}

\section{Conclusions}
\label{sec:conclusions}

We compare 1D evolutionary models of massive and very massive stars from five independent simulations. \revision{Focusing on near-solar composition, we find that the predictions from different codes can differ from each other by more than 1000\Rsun{} in terms of maximum radial expansion achieved by the stars, by
$\approx$18 percent in terms of ionizing radiation,
and about 20\Msun{} in terms of the stellar remnant mass. }
The differences in the evolution of massive stars can arise due to physical inputs like chemical abundances, mass-loss rates, and internal mixing properties. However, very massive stars, that is stars with initial masses 40\Msun{} or more, show a larger difference in evolutionary properties compared to lower mass stars. For these stars, the differences in the evolution can be largely attributed to the numerical treatment of the models when the Eddington-limit is exceeded in their low-density envelopes. 

The different methods used by 1D codes to compute the evolution of massive stars beyond density inversions (or to avoid the inversions) can modify the radius and temperature of the star, and can therefore affect the mass-loss rates.
A phenomenological justification for the mass loss enhancement comes from the fact that there are stars observed with extremely high, episodical mass loss, i.e. luminous blue variables \citep{Bestenlehner2014,Sarkisyan:2020}. 
However, other studies, such as the recent measurement of a approximately 20\Msun{} black hole in the Galactic black hole high-mass X-ray binary Cyg X-1  \citep[][]{Miller-Jones:2021}, suggest that the mass-loss rates for massive stars at near-solar metallicity may be lower than usually assumed in the 1D stellar models \citep{Neijssel2021}. 
% Several authors have tried to address the issue by computing new mass-loss rates \citep[e.g.,][]{Vink:2011, Grafener:2012} for stars near the Eddington-limit.
The exact nature of wind mass loss for very massive stars remains disputed \citep{Smith:2015}. Moreover, variation in remnants masses in Figure~\ref{fig:remnant} shows other uncertainties in massive star evolution can lead to differences at least as large as variations in mass-loss rates, which could also easily explain the formation of a 20\Msun{} black hole in Cyg X-1 in the Galaxy.
 
% However, other studies suggest that such high mass-loss rates might not be feasible for single stars, and perhaps other processes such as binary interactions might remove surface layers \citep{Smith:2015}.

None of the solutions that the BoOST, Geneva, MIST and PARSEC models employ can currently be established as better than the others. 
In each case they have been designed to address numerical issues in 1D stellar evolution. 
However, the interplay of these solutions with mass-loss rates and convection further adds to the uncertainties in massive stellar evolution. \revision{Therefore, a systematic study to untangle the effect of the treatment of the Eddington-limit from other physical assumptions has been conducted in a companion paper \citep{Agrawal2021b}.} 
% \footnote{Note to the referee: the paper has been submitted already and can be found as a pre-print with the title `A systematic study of super-Eddington envelopes in massive stars'.}

% (\todo{add reference to chapter3}).
% \poojan{Should I add reference to the other paper (Agrawal et al. in prep) here?}

In the case of BPASS the stellar models evolve without requiring any numerical enhancement. Whether this is a result of using a non-Lagrangian mesh (the `Eggletonian' mesh, which is more adaptive to changes in stellar structure) or if this is an artifact of bigger time steps (that helps stars skip problematic short-lived phases of evolution) is currently not known. A separate study to explore the effect of the `Lagrangian' versus the `Eggletonian' mesh structure for massive stars (similar to \citealt{Stancliffe:2004} study for low and intermediate mass stars) is highly desirable.

In conclusion, it is crucial to be aware of the uncertainties resulting from numerical methods whenever the evolutionary model sequences of massive stars are applied in any scientific project, such as gravitational-wave event rate predictions or star-formation and feedback studies. 
% Further physical assumptions may also play some, albeit probably minor, role in these differences. 

% After discussing the various methods that stellar evolution codes use to overcome the associated numerical convergence issues,
%We highlight the importance of proximity to the Eddington-limit in the stellar envelope and the various methods stellar evolution codes employ to overcome the associated numerical convergence issues. 
% inherent in the modelling process of massive stars

We only focus on massive stars as isolated single stars in this work.
However, there is mounting evidence that massive stars are formed as binaries or triples, thus treating them as single stars might not be correct \citep[e.g.,][]{Klencki:2020,Laplace2021}. Several studies have shown that binarity can heavily influence the lives of massive stars through mass and angular momentum transfer \revision{\citep{deMink:2009b,Marchant:2016,Eldridge2017BPASS}} and can therefore help in avoiding density and pressure inversion regions in stellar envelopes \citep{Shenar:2020}.

We also limit our study to massive stars at near-solar metallicity, where due to high opacity, the numerical instabilities related to the proximity to the Eddington-limit are maximum. Since opacity decreases with metallicity, opacity peaks become less prominent at lower metallicity. Nevertheless, stars with low-metal content also reach the Eddington-limit, although at higher initial masses \citep{Sanyal:2015,Sanyal:2017}. 
While progenitors of currently detectable gravitational-wave (GW) sources may have been born in the early Universe where the metal content is sub-solar \citep[cf.][]{Santoliquido:2021}, high star-formation rates at near solar metallicities offer a fertile ground for the formation of more GW sources, although less massive compared to sub-solar metallicity \citep{Neijssel:2019}. 
As such, there is good motivation for studying the behaviour and reliability of massive star models across a wide range of metallicities \citep[][]{Agrawal2021b}. 
% \textbf{We explore the effect of the Eddington-limit at sub-solar metallicities in forthcoming papers.}

% chapter \todo{add chapter} and chapter \todo{add chapter}.

% \jarrod{It is not really clear what point you are trying to make here, e.g. that it is not so important to look at low metallicity model behaviour or that it is something that should be done.}
% Still, since the evolution of massive stars is even less constrained at lower metallicities \citep[obtaining statistically meaningful samples of high-mass stars outside the Milky Way is a challenge, cf.][]{Garcia:2019b}, there are a number of additional uncertainties inherent in their evolution \citep{Szecsi:2015,Kubatova:2019,Sander:2020}. 

Collecting observational data as well as improvements in 3D and hydrodynamical modelling will help us better constrain the models of massive stars in the future. Until then, however, we urge the broader community to treat any set of stellar models with caution. 
Ideally, one would implement all available simulations as input into any given astrophysical study, and test the outcome also in terms of stellar evolution related uncertainties. With tools such as METISSE \citep{Agrawal:2020} and SEVN \citep{Spera:2019}, this task is becoming feasible.

\section*{Acknowledgements}

DSz has been supported by the Alexander von Humboldt Foundation. PA, SS and JH acknowledges the support from the Australian Research Council Centre of Excellence for Gravitational Wave Discovery (OzGrav). We also thank Stefanie Walch-Gassner, Debashis Sanyal and Ross Church for useful comments and discussions. \revision{We are also grateful to the referee, 
Cyril Georgy for their suggestions which greatly improved the work.}

\section*{Data Availability}

All the stellar models used in this work are publicly available: \\
PARSEC: \href{https://people.sissa.it/~sbressan/parsec.html}{people.sissa.it/$\sim$sbressan/parsec.html},\\
MIST: \href{http://waps.cfa.harvard.edu/MIST/}{waps.cfa.harvard.edu/MIST/},\\
Geneva: \href{https://obswww.unige.ch/Research/evol/tables_grids2011/Z014/}{obswww.unige.ch/Research/evol/tables\_grids2011/Z014},\\
BPASS: \href{https://bpass.auckland.ac.nz/9.html}{bpass.auckland.ac.nz/9.html},\\
BoOST: \href{http://boost.asu.cas.cz}{boost.asu.cas.cz}.

%%%%%%%%%%%%%%%%%%%% REFERENCES %%%%%%%%%%%%%%%%%%
\bibliographystyle{mnras}
\bibliography{references} 
%%%%%%%%%%%%%%%%%%%%%%%%%%%%%%%%%%%%%%%%%%%%%%%%%%

% \appendix

% Don't change these lines
\bsp	% typesetting comment
\label{lastpage}
\end{document}